\documentclass[aps,prl,twocolumn,groupedaddress,showpacs,preprintnumbers]{revtex4}

\def\be{\begin{equation}}
\def\ee{\end{equation}}
\def\nn{\nonumber}
\def\bea{\begin{eqnarray}}
\def\eea{\end{eqnarray}}
\def\gsim{\ \rlap{\raise 2pt\hbox{$>$}}{\lower 2pt \hbox{$\sim$}}\ }
\def\lsim{\ \rlap{\raise 2pt\hbox{$<$}}{\lower 2pt \hbox{$\sim$}}\ }
\def\dslash{\kern-4pt \not{\hbox{\kern-2pt $\partial$}}}
\def\pslash{\not{\hbox{\kern-2pt p}}}
\def\p{{\bf p}}

\def\x{{\bf x}}
\def\xp{{\bf x'}}

\def\e{{\rm e}}

\begin{document}

\preprint{FERMILAB-Pub-03/056-T}

\title{Homeotic supersymmetry}


\author{Gabriela Barenboim}
\email[]{gabriela@fnal.gov}

\author{Joseph Lykken}
\email[]{lykken@fnal.gov}
\affiliation{Fermi National Accelerator Laboratory\\
P.O. Box 500, Batavia, IL 60510 USA}



\begin{abstract}
It is commonly believed that unbroken supersymmetry (SUSY) implies that
all members of a supermultiplet have the same mass. We demonstrate
that this is not true, by exhibiting a simple counterexample. We employ
the formalism of homeotic fermions, in a simple model where $CPT$
conjugate fermions have different masses. This model can be supersymmetrized
to a hypermultiplet of fields which form a representation of the
conventional $N$$=$$2$ SUSY algebra. Nevertheless, $CPT$ conjugate states
in this hypermultiplet have different masses. These surprising results do not
violate either the $CPT$ theorem nor the Haag-Lopusza\'nski-Sohnius theorem.
\end{abstract}


\maketitle


\section{Introduction}
Supersymmetries are the only possible extensions
of the four dimensional
Poincar\'e invariance observed ubiquitously in particle interactions
~\cite{Haag:1975qh,Coleman:1967ad}.
Supersymmetry (SUSY) plays a fundamental role in string theory, and there
are many strong phenomenological motivations for believing that
supersymmetry is realized in nature, in spontaneously broken form.
On the other hand, no superpartner particles have yet been observed,
and spontaneously broken supersymmetry makes a prediction for
the cosmological vacuum energy density which is too large by at least 60
orders of magnitude. Thus it is important to push the boundaries
of our fundamental understanding of supersymmetry, and more especially
to look for novel ways of expressing supersymmetry in physical systems.

It is widely believed that unbroken supersymmetry implies that
all members of a supermultiplet have the same mass. For example
Sohnius' authoritative review article states explicitly that
``supersymmetry must be broken in nature where elementary particles
do not come in mass-degenerate multiplets''~\cite{Sohnius:1985qm}.
This belief is based upon the strong constraint that any
conserved supercharge $Q$ must commute with the 4-momentum operator
$P_{\mu}$, which in turn implies the O'Raifeartaigh theorem:
\be
[Q,P_{\mu}P^{\mu}] = 0 \quad .
\ee

In spite of these facts, we demonstrate in
this letter that exact supersymmetry does {\it not} always
imply mass-degenerate multiplets. We employ
the formalism of homeotic fermions~\cite{Barenboim:2002tz}, 
developed previously by
us as a toy model for $CPT$ violation in the
neutrino sector
~\cite{Barenboim:2001ac,Barenboim:2002rv,Barenboim:2002hx,Barenboim:2002ah,
Barenboim:2003jm}.
Our starting point is an extremely simple
model where $CPT$
conjugate fermions have different masses. This model can be supersymmetrized
to a hypermultiplet of fields which form a representation of the
conventional $N$$=$$2$ SUSY algebra. Nevertheless, $CPT$ conjugate states
in this hypermultiplet have different masses. Each fermion state is still
mass-degenerate with a boson state, however the hypermultiplet is {\it not}
reducible to a pair of mass-degenerate supermultiplets.
These surprising results do not
violate either the $CPT$ theorem~\cite{Streater:1989vi}
nor the Haag-Lopusza\'nski-Sohnius theorem~\cite{Haag:1975qh}.

\section{Homeotic fermions}

Homeotic fermions, like Dirac fermions, are denoted by 4-component
complex spinor fields $\psi(x)$. In the free homeotic theory
the Fourier-transformed fermion
fields obey the equation of motion:
\be
(\pslash - m\epsilon(p_0))\psi(p) = 0,
\label{heom}
\ee
which differs from the Dirac equation by the presence of
$\epsilon(p_0)$, the sign function of
delta calculus. Solutions of (\ref{heom}) are solutions
of the Klein-Gordon equation; the homeotic theory can be regarded
as ``the other square root'' of the Klein-Gordon equation.
The homeotic case is usually neglected because the equation of motion
is nonlocal in position space:
\be
i\dslash \psi(t,\x) = -{im\over\pi}\,{\bf P}\int dt^{\prime}\,
{1\over t - t^{\prime}} \,\psi(t^{\prime},\x) \;,
\ee
where {\bf P} denotes the principal value integral, which
we will now assume throughout. Despite this nonlocality, it
was demonstrated in \cite{Barenboim:2002tz} that causal interacting
homeotic field theories exist, in the sense of having a well-defined
perturbative S-matrix.

As noted in \cite{Barenboim:2002tz}, the combination of a homeotic
mass term with a Dirac mass term violates $CPT$. Consider the
simple free theory defined by the lagrangian
\be
\int d^3x\, \bar{\psi} (i\dslash - m_d) \psi
+ {im_h\over \pi} \int
{d^3xdt'\over t-t'} \left( \bar{\psi}(t) \psi(t') -  \bar{\psi}(t') \psi(t)
\right) \, .
\label{homaction}
\ee
If we define the $CPT$ operator such that the Dirac mass term
is $CPT$ even, then the homeotic mass term is $CPT$ odd, and vice-versa.
$CPT$ conjugate spinors in this theory have mass-squared eigenvalues
$(m_d\pm m_h)^2$.

\section{The $N$$=$$2$ SUSY algebra}

We would now like to extend this theory of 4-component complex spinors
to a supermultiplet of fields which furnish a representation of the
standard $N$$=$$2$ SUSY algebra. In the 4-component notation of
Sohnius~\cite{Sohnius:1985qm} the relevant parts of the $N$$=$$2$
algebra are:
\bea
\{ Q^i, \overline{Q}_j \} &=& 2\delta^i_j \gamma^{\mu}P_{\mu}
+ 2i\delta^i_j Z \, ,\cr
[ Q_i , P_{\mu} ] &=& 0 \, ,
\label{ntwoalgebra}
\eea
where the $Q_i$, $i$$=$$1$, 2, are symplectic Majorana spinor supercharges,
and $Z$ is the (antihermitian) central charge operator, which commutes
with all of the other generators of the algebra.
The supercharges form a doublet under the $SU(2)$ $R$ symmetry of
$N$$=$$2$ SUSY; the index $i$ is raised and lowered with the two-dimensional
Levi-Civita tensors $\epsilon_{ij}$$=$$\epsilon^{ij}$. In our derivation
we will need a number of identities for bilinears of symplectic
Majorana spinors:
\bea
\bar{\zeta}^i\eta_i 
&=& -(\bar{\zeta}^i\eta_i)^{\dagger}
= \bar{\eta}_i\zeta^i = -\bar{\zeta}_i\eta^i \,, \cr
\bar{\zeta}^i\gamma^{\mu}\eta_i 
&=& -(\bar{\zeta}^i\gamma^{\mu}\eta_i)^{\dagger}
= \bar{\eta}_i\gamma^{\mu}\zeta^i = -\bar{\zeta}_i\gamma^{\mu}\eta^i \,, \cr
\bar{\zeta}_i\eta^j - \bar{\eta}_i\zeta^j
&=& \delta_i^j\bar{\zeta}_k\eta^k \,,\cr
\bar{\zeta}_i\gamma^{\mu}\eta^j - \bar{\eta}_i\gamma^{\mu}\zeta^j
&=& \delta_i^j\bar{\zeta}_k\gamma^{\mu}\eta^k \,.
\label{bilinears}
\eea

Let $A_i(x)$ denote a doublet of complex scalar fields, and $\psi(x)$
a 4-component complex fermion field. We want these fields to form an
$N$$=$$2$ hypermultiplet, {\it i.e.}, to furnish a representation of
the algebra (\ref{ntwoalgebra}). The fundamental relation between
the fields is
\be
[Q^i,A_j(x)] = -i\sqrt{2}\delta^i_j\,\psi(x) \; .
\label{basicapsi}
\ee
one additional input is required:
\be
[Z,A_i(x)] = F_i \; ,
\label{basicf}
\ee
where $F_i(x)$ is a doublet of complex bosonic auxiliary fields;
we will specify the precise form of $F_i(x)$ later.

Expressions (\ref{basicapsi}) and (\ref{basicf}), together with
the algebra (\ref{ntwoalgebra}), the Jacobi identities,
and the identities (\ref{bilinears}), now
imply:
\bea
\{\overline{Q}_i,\psi\} &=& \sqrt{2}i\gamma^{\mu}[P_{\mu},A_i]
- \sqrt{2}F_i \, ,\cr
[Z,\psi ] &=& -i\gamma^{\mu}[P_{\mu},\psi ] = \;{\dslash}\psi \, ,\cr
[Q^i,F_j] &=& -i\sqrt{2}\delta^i_j[Z,\psi ] \, ,\cr
[Z,F_i] &=& -[P^{\mu},[P_{\mu},A_i]] = 
\partial_{\mu}\partial^{\mu}A_i \, .
\label{jacobi}
\eea

These relations define a set of hypermultiplet
fields $A_i(x)$, $\psi(x)$, and
$F_i(x)$ which form a representation of $N$$=$$2$ SUSY.
These relations determine uniquely the equations of motion for
the dynamical fields $A_i(x)$ and $\psi(x)$, once we specify the
auxiliary fields $F_i(x)$. If {\it e.g.} we write
\be
F_i(x) = -im_dA_i \; ,
\ee
then (\ref{basicf}) combined with (\ref{jacobi}) implies
that the $A_i(x)$ satisfy the Klein-Gordon equation, while
(\ref{basicapsi}) combined with (\ref{jacobi}) implies that
$\psi(x)$ obeys the Dirac equation.

\section{Simple example of homeotic supersymmetry}

We obtain the simplest example of homeotic supersymmetry by
specifying the auxiliary fields as follows:
\be
F_i(x) = -im_dA_i(x) - {m_h\over\pi}\int{dt'\over t-t'}A_i(t',\x) \,.
\label{eomone}
\ee 
Applying the relations (\ref{basicapsi}-\ref{jacobi}), we determine
the equations of motion to be:
\bea
&\displaystyle{\partial_{\mu}\partial^{\mu}A_i
+(m_d^2+m_h^2)A_i -{2im_dm_h\over\pi}\int {dt'\over t-t'}A_i(t',\x) = 0
,}\cr
~~\cr
&\displaystyle{i\dslash\psi - m_d\psi +{im\over\pi}\int {dt'\over t-t'} 
\psi(t',\x) =0
.}
\label{eomtwo}
\eea
The action which reproduces (\ref{eomone}), (\ref{eomtwo}) is
given by
\be
S = \int dt \left( L_{kin} + L_d + L_h + L_h^{\dagger}\right) \; ,
\label{ouraction}
\ee
where
\bea
L_{kin} &=& \int d^3x \left[
\partial_{\mu}A^{\dagger i}\partial^{\mu}A_i
+F^{\dagger i}F_i
+ i\bar{\psi}\dslash\psi \right] \, ,\cr
L_d &=& -m_d \int d^3 x \left[
iA^{\dagger i}F_i - iF^{\dagger i}A_i + \bar{\psi}\psi \right] \, ,
\eea
\vspace{-0.7cm}
\begin{eqnarray*}
L_h = {im_h\over \pi}\int {d^3xdt'\over t-t'} \left[
iA^{\dagger i}(t)F_i(t') - iF^{\dagger i}(t)A_i(t')
+ \bar{\psi}(t)\psi(t') \right] \, .\nn
\end{eqnarray*}

It is easy to check that all four pieces
of the action are separately invariant under the
standard $N$$=$$2$ SUSY transformations of the fields:
\bea
\delta A_i &=& \sqrt{2}\bar{\zeta}_i\psi \; ,\cr
\delta\psi &=& -i\sqrt{2}F_i -i\sqrt{2}\gamma^{\mu}\zeta^i
\partial_{\mu}A_i \; ,\cr
\delta F_i &=& \sqrt{2}\bar{\zeta}_i\dslash\psi \; .
\label{susytrans}
\eea

We can proceed further to construct the conserved supercurrent
in terms of the component fields. However the usual Noether procedure does
not yield a conserved current; this is a generic feature of nonlocal
field theories, and was noted in our previous work with regard to
the fermion number current of the homeotic fermion theory.
Let us first review that case, in which
we employed a trick of Pauli's construct the conserved
fermion number current:
\bea
J^{\mu} &=& \bar{\psi}\gamma^{\mu}\psi \cr
&+& \delta^{\mu 0}{m_h\over\pi}\int^t dt'\int {dt''\over t'-t''}
\left[ \bar{\psi}(t')\psi(t'')+\bar{\psi}(t'')\psi(t')
\right] \, . \nn
\label{fncurrent}
\eea
It is easily seen using the
fermion equation of motion (\ref{eomtwo})
that $J^{\mu}(x)$ is conserved on-shell. Despite its ugly form
in position space, $J^{\mu}(x)$ reduces to the usual number
current in the creation/annihilation Fock basis.

\begin{widetext}
In a supersymmetric theory the conserved bosonic current
$J^{\mu}(x)$ must belong to a supermultiplet of conserved currents.
In particular, we can immediately obtain an expression for the
supercurrent $j^{\mu i}$, by applying the supersymmetry transformations
(\ref{susytrans}) to the component fields in
(\ref{fncurrent}). The result is:
\be
j^{\mu i}(x)/\sqrt{2} = \gamma^{\mu}\psi F^{\dagger i} + \gamma^{\nu}
\gamma^{\mu}\psi\partial_{\nu}A^{\dagger i} 
+\delta^{\mu 0}{m_h\over\pi}\int^t dt'
\int {dt''\over t'-t''} \left[
\psi(t'')F^{\dagger i}(t') + \gamma^{\nu}\psi(t'')\partial_{\nu}
A^{\dagger i}(t') + (t'\leftrightarrow t'')\right] \, .
\label{supercurrent}
\ee
\end{widetext}
Using the equations of motion (\ref{eomone}), (\ref{eomtwo}), one
finds that $j^{\mu i}(x)$ is conserved on-shell. It thus represents
the supercurrent modulo possible ``improvement'' terms~\cite{Sohnius:1985qm}.

\section{Mass spectrum}

Both the bosonic and fermionic parts of the action
(\ref{ouraction}) violate $CPT$. Let us focus first on the
bosonic sector. We can expand the fields $A_i(x)$ in 
positive and negative frequency plane
wave solutions of the equations of motion (\ref{eomtwo}):
\be
A_i = \int{d^3p\over (2\pi )^3}(
{a_{\p i}\over\sqrt{2\omega_+}}\e^{-i\omega_+ t+i\p\cdot\x}
+ {b_{\p i}^{\dagger}\over\sqrt{2\omega_-}}
\e^{i\omega_- t-i\p\cdot\x}),
\ee
where
\be
\omega_{\pm} \equiv \sqrt{\p^2 + (m_d\pm m_h)^2} \; .
\label{omegapm}
\ee
We quantize the theory by assuming that $a_{\p i}$, $b_{\p i}$
satisfy the commutation relations of creation/annihilation operators.
It follows that the general commutator of $A_i(x)$ with its
conjugate $\Pi^j(x)$ is given by
\bea
[ A_i(x), \Pi^j(x') ] =
{i\over 2}\delta_i^j\int {d^3p\over (2\pi )^3}&\cr
\times( \e^{-i\omega_+(t-t')+i\p\cdot (\x -\xp )}
+& \e^{i\omega_-(t-t')-i\p\cdot (\x -\xp )}).
\nn
\label{gencom}
\eea
Thus the equal-time commutator is canonical, but the general
commutator is not. Using (\ref{gencom}) we can now verify that
the supercharge extracted from (\ref{supercurrent}) satisfies
(\ref{basicapsi}).

Another novel feature appears when we construct the bosonic part
of the hamiltonian in terms of $a_{\p i}$, $b_{\p i}$. The
canonical hamiltonian is not diagonalized in the basis defined
by (\ref{omegapm}); it is instead diagonalized in the basis
defined by
\be
A_i = \int{d^3p\over (2\pi )^3}(
{a_{\p i}\over\sqrt{2\omega_+}}\e^{-i\omega t+i\p\cdot\x}
+ {b_{\p i}^{\dagger}\over\sqrt{2\omega_-}}
\e^{i\omega t-i\p\cdot\x}),
\ee
where
\be
\omega = \sqrt{\p^2 + m_d^2} \; .
\label{omegad}
\ee
In this basis the bosonic hamiltonian is
\be
H_b = \int d^3x \left[
\omega_+ a_{\p}^{\dagger i}a_{\p i} + \omega_- b_{\p}^{\dagger i}b_{\p i}
\right]\; ,
\ee
showing that the $CPT$ conjugate single particle states have
a mass-squared splitting equal to $\vert 4m_dm_h\vert$.

A similar analysis for the fermions diagonalizes the
fermionic part of the hamiltonian in terms of the anticommuting
Fock operators 
$\tilde{a}_{\p s}$, $\tilde{b}_{\p s}$, where $s$ is the spin label:
\be
H_f = \int d^3x \left[
\omega_+ 
\tilde{a}_{\p s}^{\dagger}
\tilde{a}_{\p s} + \omega_- 
\tilde{b}_{\p s}^{\dagger}
\tilde{b}_{\p s}
\right]\; .
\ee
Again the $CPT$ conjugate states have a mass-squared
splitting equal to $\vert 4m_dm_h\vert$.

\section{Comments}

It would be interesting to extend the above construction to
produce an interacting theory.
A conventional $N$$=$$2$ hypermultiplet can interact with
an $N$$=$$2$ vector multiplet, or can have self-interactions
describing a nonlinear sigma model~\cite{Gates:1983nr}.
For the homeotic case neither extension appears entirely
straightforward.

The homeotic $N$$=$$2$ hypermultiplet
has 8 real on-shell degrees of freedom. We have just seen
that half of these describe a boson-fermion pair with
mass $\vert m_d + m_h\vert$, while the other half describe
a boson-fermion pair with mass $\vert m_d - m_h\vert$.

Since an ordinary $N$$=$$2$ hypermultiplet can be split into
two $N$$=$$1$ chiral multiplets, it is important to ask
whether our homeotic $N$$=$$2$ hypermultiplet is reducible into
two $N$$=$$1$ multiplets. The answer is no. This can be seen
by imposing the Majorana condition on the fermions in
(\ref{homaction}), and observing that the homeotic mass term
then vanishes identically. Alternatively, one notes that
the usual decomposition of the hypermultiplet into chiral
multiplets can be written in the Fock basis as
\be
a_{\p i},\,b_{\p i},\,
\tilde{a}_{\p s},\,\tilde{b}_{\p s} \rightarrow
(a_{\p\pm},\, b_{\p\mp},\, \tilde{a}_{\p\pm},\,
\tilde{b}_{\p\mp}),\; ;
\ee
obviously in the homeotic case this would mix operators with
different dispersion relations.

The irreducibility of the homeotic $N$$=$$2$ hypermultiplet is
in fact very analogous to the irreducibility of the
ordinary $N$$=$$1$ chiral multiplet. In this case one finds
the 4 on-shell degrees of freedom consist of two $CPT$ conjugate
pairs; each pair has one boson and one fermion state, related by
supersymmetry. However it is not possible to reduce the multiplet,
due to the non-existence of Majorana-Weyl spinors
in four dimensions~\cite{Lykken:1996xt}.
This is the analog of the non-existence of homeotic Majorana spinors
in four dimensions.

As a parting remark, let's inquire how one might attempt to
contruct a supermultiplet in which bose-fermi degeneracy is
violated. At the level of the on-shell hamiltonian, this does
not appear to be particularly difficult. Let $H_0(m)$ be a
mass term with mass parameter $m$ for a
free supersymmetric hamiltonian containing two species of fermions,
and construct a new hamiltonian
defined by
\be
H_1  = H_0(m) + (-1)^{F_1+F_2}H_0(m') \quad ,
\ee
where $F_1$ and $F_2$ are the fermion number operators for the
two species of fermions. 
Clearly the single particle eigenstates of $H_1$
have different masses for bosons and fermions:
the bosons have mass $m+m'$ while the fermions have mass $m-m'$.
It is easy to see that, acting
on single particle states:
\be
\{ Q,(-1)^{F_1} \} = \{ Q,(-1)^{F_2} \} = 0 \quad,
\ee
from which it follows that, acting on single particle states:
\be
[ Q,H_0 ] = 0 \Rightarrow [ Q,H_1 ] = 0 \; ,
\ee
and thus supersymmetry is unbroken.
The challenge, of course, is to realize such a scheme in
field theory.



\begin{acknowledgments}
We would like to acknowledge helpful discussions with 
Bill Bardeen, Piyush Kumar, Chris Quigg, and Erick Weinberg.
This research was supported by the U.S.~Department of Energy
Grants DE-AC02-76CHO3000 and DE-FG02-90ER40560.
\end{acknowledgments}

\bibliography{hrefs}

\end{document}